\documentclass[preprint2]{aastex6}
\newcommand{\dbr}{\partial_t B_r}
\newcommand{\dbz}{\partial_t B_z}
\newcommand{\Ep}{{\mathbf E}_\perp}
\newcommand{\Epo}{{\mathbf E}_{\perp 0}}
\newcommand{\bb}{{\mathbf b}}
\newcommand{\nb}{{\mathbf n}}
\newcommand{\vb}{{\mathbf v}}
\newcommand{\xb}{{\mathbf x}}
\newcommand{\Bb}{{\mathbf B}}
\newcommand{\Eb}{{\mathbf E}}
\newcommand{\evx}{{\mathbf e}_x}

\newcommand{\evz}{{\mathbf e}_z}
\newcommand{\evt}{{\mathbf e}_\theta}
\newcommand{\evp}{{\mathbf e}_\phi}
\newcommand{\evr}{{\mathbf e}_r}
\newcommand{\dS}{\,\mathrm{d}S}

\renewcommand{\edit}[1]{}

\fullcollaborationName{The Friends of AASTeX Collaboration}

\begin{document}


\title{Sparse Reconstruction of Electric FIelds from Radial Magnetic Data}

%% Use \author, \affil, plus the \and command to format author and affiliation 
%% information.  If done correctly the peer review system will be able to
%% automatically put the author and affiliation information from the manuscript
%% and save the corresponding author the trouble of entering it by hand.
%%
%% The \affil should be used to document primary affiliations and the
%% \altaffil should be used for secondary affiliations, titles, or email.

%% Authors with the same affiliation can be grouped in a single
\author{Anthony R. Yeates}
\affil{Department of Mathematical Sciences \\
Durham University \\
Durham, DH1 3LE, UK}

\begin{abstract}

Accurate estimates of the horizontal electric field on the Sun's visible surface are important not only for estimating the Poynting flux of magnetic energy into the corona but also for driving time-dependent magnetohydrodynamic models of the corona. In this paper, a method is developed for  estimating the horizontal \edit{electric} field from a sequence of radial-component magnetic field maps. This problem of inverting Faraday's law has no unique solution. Unfortunately, the simplest solution (a divergence-free electric field) is not realistically localized in regions of non-zero magnetic field, as would be expected from Ohm's law. Our new method generates instead a localized solution, using a basis pursuit algorithm to find a sparse solution for the electric field. The method is shown to perform well on test cases where the input magnetic maps are flux balanced, in both Cartesian and spherical geometries. However, we show that if the input maps have a significant imbalance of flux -- usually arising from data assimilation  -- then it is not possible to find a localized, realistic, electric field solution. This is the main obstacle to driving coronal models from time sequences of solar surface magnetic maps.
\end{abstract}




%%%%%%%%%%%%%%%%%%%%%%%
%%%%%%%%%%%%%%%%%%%%%%%

\section{Introduction} \label{sec:intro}

Magneto-hydrodynamic (MHD) simulations of the magnetic field in the Sun's corona are typically driven by an imposed evolution on the solar surface. Since the magnetic field $\Bb$ evolves according to Faraday's law
\begin{equation}
\partial_t\Bb = -\nabla\times\Eb,
\label{eqn:faraday}
\end{equation}
the required boundary condition is the horizontal electric field $\Ep = E_\theta(\theta,\phi,t)\evt + E_\phi(\theta,\phi,t)\evp$, written here in spherical polar coordinates. Unfortunately, this electric field cannot be observed directly, but must be reconstructed from other observations. If we assume that the plasma obeys the ideal Ohm's law $\Eb = -\vb\times\Bb$ then $\Eb_\perp$ can, in principle, be computed from vector observations of $\Bb$ and of the plasma velocity $\vb$. \edit{\citet{fisher2010}, \citet{fisher2012} and \citet{kazachenko2014} have developed the most comprehensive method of estimating electric fields using both vector magnetograms and Doppler velocity measurements, and have successfully applied this to observations of active region 11158 \citep{kazachenko2015}.}

In practice, however, such observations are not routinely available for the full solar surface. Many authors have therefore opted to estimate $\Ep$ purely by inverting Equation (\ref{eqn:faraday}), typically using only a time sequence of $B_r(\theta,\phi,t)$ data \citep{mikic1999, wu2006, mackay2011, cheung2012, yang2012}. In this case, only the radial component of (\ref{eqn:faraday}) is used,
\begin{equation}
\dbr = -\evr\cdot\nabla\times\Ep.
\label{eqn:faraday1}
\end{equation}
The present paper seeks to address one of the potential pitfalls of this technique. 

The fundamental problem of electric field inversion from Faraday's law (\ref{eqn:faraday}) is lack of uniqueness. If we write $\Ep$ as a Helmholtz decomposition
\begin{equation}
\Ep = -\nabla\times\big(\Phi\evr\big) - \nabla_\perp\Psi,
\label{eqn:ep}
\end{equation}
then $\nabla\times\Ep$ depends only on the potential $\Phi(\theta,\phi,t)$, and is independent of the choice of the second potential $\Psi(\theta,\phi,t)$. Note that our notation interchanges $\Phi$ and $\Psi$ compared to \citet{mikic1999}. \edit{In the notation of \citet{kazachenko2014}, our $\Phi$ and $\Psi$ correspond  to $\dot{\cal B}$ and $\dot{\cal J}$, respectively.} Given that $\dbr$ is independent of $\Psi$, the simplest way to obtain a solution consistent with the given $B_r$ data is to set $\Psi\equiv 0$. Then (\ref{eqn:faraday1}) becomes the Poisson equation
\begin{equation}
-\nabla^2\Phi = \dbr,
\label{eqn:poisson}
\end{equation}
which may be solved uniquely for $\Phi$ given appropriate boundary conditions. We call the resulting $\Ep$ the \emph{inductive} solution for the electric field, and denote it by $\Epo$. This solution has been used to drive coronal models in numerous studies \citep[e.g.,][]{mikic1999,amari2003, mackay2011, cheung2012, gibb2014, yang2012, feng2015}. Unfortunately, there is good reason to believe that the non-inductive potential $\Psi$ is non-negligible on the real solar surface. \edit{For example, \citet{fisher2010} and \cite{kazachenko2014} have shown that a non-inductive term is required to correctly reconstruct the electric field from a numerical MHD simulation.}

\begin{figure*}
\includegraphics[width=\textwidth]{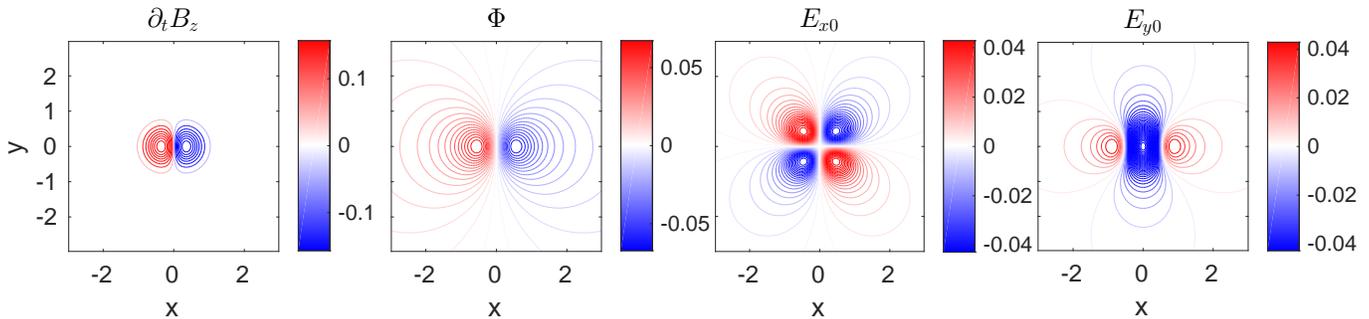}
\caption{The simple solution in Equation (\ref{eqn:anal}), showing $\dbz$, the inductive potential $\Phi$, and the resulting $E_{x0}$, $E_{y0}$. The color scales for $\dbz$, $E_{x0}$, and $E_{y0}$ are clipped at one fifth of their maximum.}
\label{fig:analytical}
\end{figure*}

One problem with the inductive solution is its lack of localization. If the real $\Ep$ satisfies Ohm's law, then it ought to vanish outside of patches of strong $B_r$. As discussed in Section \ref{sec:inductive}, this is not the case with the inductive solution; the main focus of this paper will be to choose $\Psi$ in order to correct this deficiency.

A second problem with the inductive solution is that it cannot detect electric fields associated with plasma flows along contours of $B_r$, since $\dbr=0$ in that case. The corresponding contribution to $\Psi$ may be added if the flows are known. For example, \citet{kazachenko2014,kazachenko2015} have made this correction using observations of plasma velocities in active regions, and the resulting $\Ep$ has been used to drive magneto-frictional simulations \citep{fisher2015}. On the other hand, \citet{weinzierl2016} have included  the contribution to $\Ep$ from a prescribed large-scale differential rotation in their global magneto-frictional simulations. In this paper, however, we will put this issue aside and focus only on the localization problem.

In Section \ref{sec:inductive}, we describe the localization problem with the inductive solution, before characterizing the inductive solution in a variational framework. This makes the link to our proposed sparse electric field solution, described in Section \ref{sec:sparse}, that is designed to restore localization. Numerical tests in both Cartesian and spherical geometry are presented in Section \ref{sec:tests}, while Section \ref{sec:limitations} considers the limitations of the sparse solution.

%%%%%%%%%%%%%%%%%%%%%%%
%%%%%%%%%%%%%%%%%%%%%%%

\section{Inductive Electric Field} \label{sec:inductive}

\subsection{The localization problem}

Since the inductive potential $\Phi$ solves the Poisson equation (\ref{eqn:poisson}), the solution may be expressed in terms of the Green's function as
\begin{eqnarray}
\Phi(x,y,t) = \frac{1}{2\pi}\int_{-\infty}^\infty\int_{-\infty}^\infty \dbz(x',y',t)\nonumber\\
\qquad\times\log\big|(x-x')^2 + (y-y')^2\big|^{1/2}\,\mathrm{d}x'\,\mathrm{d}y'.
\end{eqnarray}
Here we have taken Cartesian coordinates and an infinite plane for simplicity, but similar solutions hold on a spherical surface. For a localized source $\dbz$, we therefore have that $\Phi\sim \log(r)$ for large distance $r$ from the source. The corresponding inductive electric field $\Epo$ therefore decays only as $r^{-1}$. In particular, it may be non-zero well outside the regions of non-zero $B_z$. 

As an explicit example, consider a bipolar distribution
\begin{equation}
\dbz = \exp\left(-\frac{(x + \rho)^2 + y^2}{\delta^2}\right) - \exp\left(-\frac{(x - \rho)^2 + y^2}{\delta^2}\right).
\end{equation} 
In this case, one may solve Equation (\ref{eqn:poisson}) exactly to obtain the closed-form solution
\begin{equation}
\Phi(x,y) = \frac{\delta^2}{2}\log\left(\frac{r_+}{r_-}\right) + \frac{\delta^2}{4}E_1\left(\frac{r_+^2}{\delta^2}\right) - \frac{\delta^2}{4}E_1\left(\frac{r_-^2}{\delta^2}\right),
\label{eqn:anal}
\end{equation}
where $r_\pm = \sqrt{(x\pm\rho)^2 + y^2}$ and $E_1(x) := \int_x^\infty\mathrm{e}^{-t}/t\,\mathrm{d}t$ is the exponential integral. The logarithmic behaviour is clear from this expression, and the corresponding electric field components are plotted in Figure \ref{fig:analytical}.

\subsection{Variational formulation}

Another way to characterise the inductive solution $\Epo$ is by the fact that it minimizes the $L_2$-norm $\|\Ep\|_2 := \left(\int_S|\Ep|^2\dS\right)^{1/2}$ among all possible solutions to (\ref{eqn:faraday1}). This is straightforward to see by inserting the expression (\ref{eqn:ep}) into the norm, which leads to
\begin{eqnarray}
\|\Ep\|_2^2 = \|\Epo\|_2^2 + \|\nabla_\perp\Psi\|_2^2 + 2\oint_{\partial D}\Psi\Epo\cdot\nb\dS.
\end{eqnarray}
On a periodic or infinite domain, the boundary term vanishes, and it follows that $\|\Ep\|_2$ is minimized by choosing $\Psi\equiv 0$. In this sense, the inductive electric field $\Epo$ is the smoothest possible $\Ep$ satisfying Faraday's law for a given $\dbz$.

\subsection{Discrete formulation} \label{sec:dis}

\begin{figure}
\centering
\includegraphics[width=\columnwidth]{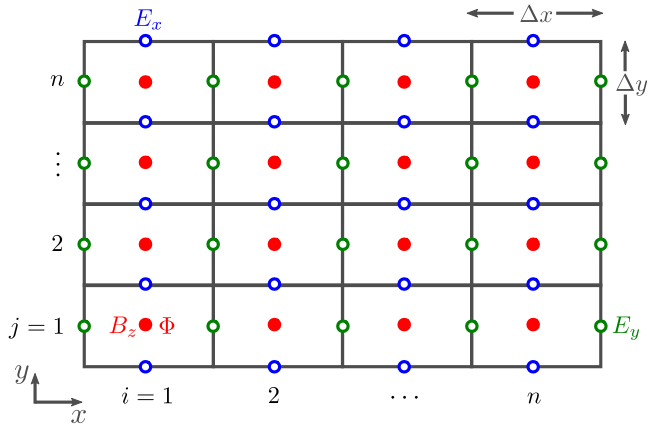}
\caption{The staggered grid, where $B_z$ and $\Phi$ are defined at cell centres and $E_x$ and $E_y$ on corresponding cell edges.}
\label{fig:grid}
\end{figure}

In practice, we work with $\Ep$ defined on a discrete numerical grid, so it is useful to formulate the analogous discrete problem. In this paper, we discretise $\Ep$ on a staggered grid \citep{yee1966}, where $E_x$ is defined on the horizontal cell edges, $E_y$ on the vertical cell edges, and $\dbz$ at the cell centers (Figure \ref{fig:grid}). Such grids are commonly used in MHD simulations. We discretise Faraday's law (\ref{eqn:faraday1}) using Stokes' theorem, so that the equation
\begin{eqnarray}
\Delta x\Delta y \frac{\partial B_z^{i,j}}{\partial t} = \Delta x E_x^{i,j+1/2} - \Delta x E_x^{i,j-1/2}\nonumber\\
 + \Delta y E_y^{i-1/2,j} - \Delta y E_y^{i+1/2,j}
\label{eqn:discrete}
\end{eqnarray}
must hold for each grid cell $i,j=1,\ldots,n$. (In more general curvilinear coordinates, the edge lengths and cell areas would be different for each cell.) This constitutes a system of linear equations for the unknowns $E_x$ and $E_y$ on each edge. However, the system is highly under-determined, since there are $2n(n+1)$ unknowns but only $n^2$ equations. This non-uniqueness of solution reflects our freedom in choosing the non-inductive component.

In the discrete case, the inductive solution may be computed by writing
\begin{eqnarray}
E_{x0}^{i,j+1/2} &= (\Phi^{i,j} - \Phi^{i,j+1})/{\Delta y},\label{eqn:edis1}
\\
E_{y0}^{i+1/2,j} &= (\Phi^{i+1,j} - \Phi^{i,j})/{\Delta x},
\label{eqn:edis2}
\end{eqnarray}
where $\Phi^{i,j}$ is defined at the cell centres. Then (\ref{eqn:discrete}) becomes
\begin{eqnarray}
\frac{\partial B_z^{i,j}}{\partial t} = \frac{1}{\Delta y^2}\Big(2\Phi^{i,j} - \Phi^{i,j+1} - \Phi^{i,j-1}\Big) \nonumber\\
+ \frac{1}{\Delta x^2}\Big(2\Phi^{i,j} - \Phi^{i-1,j} -\Phi^{i+1,j} \Big),
\label{eqn:pdis}
\end{eqnarray}
which is simply the standard 5-point stencil for the Poisson equation. In the examples below, we solve this using a standard fast-Poisson solver \citep{press1992}.

However, it is instructive to think of the system of equations (\ref{eqn:discrete}) in the more abstract form $A\xb=\bb$, where $\xb$ is the vector of unknowns ($E_x$, $E_y$), $\bb$ is the vector of $\dbz$ values, and $A$ is the $n^2\times 2n(n+1)$ matrix corresponding to equations (\ref{eqn:discrete}). The inductive solution is then the solution to $A\xb=\bb$ that minimizes the discrete $\ell_2$-norm $\|\xb\|^2_2:=\sum_i x_i^2$. In other words, it is the (unique) least-squares solution to this under-determined system of equations. As such, the solution may be written in terms of the Moore-Penrose pseudo-inverse as $\xb = A^\top(AA^\top)^{-1}\bb$, although in practice it is much more efficient to use the fast-Poisson solver. Nevertheless, viewing the inductive solution as the $\ell_2$-minimum makes the connection with the sparse solution that we will describe in Section \ref{sec:sparse}. The sparse solution is found by minimizing a different discrete norm of $\xb$.

%%%%%%%%%%%%%%%%%%%%%%%
%%%%%%%%%%%%%%%%%%%%%%%
\section{Sparse Electric Field} \label{sec:sparse}

Our idea is to find a sparse solution for $\xb$ (i.e., $E_x$, $E_y$) that minimizes the number of non-zero values. This should be more localized than the inductive solution $\Epo$. Since it will differ from $\Epo$, this new solution will have a non-zero potential $\Psi$. However, we will work directly with the system $A\xb=\bb$ (Equation \ref{eqn:discrete}), rather than solving for $\Psi$ (or $\Phi$) explicitly.

Instead of minimizing the $\ell_2$-norm of $\xb$, as in the inductive solution, we propose to minimize the $\ell_1$-norm. In other words, minimize
\begin{equation}
\|\xb\|_1 := \sum_i|x_i| \quad \textrm{subject to} \quad A\xb=\bb.
\label{eqn:l1}
\end{equation}
In the optimization literature, problem (\ref{eqn:l1}) is known as basis pursuit \citep{chen2001}. It is used in a wide variety of fields, and has recently been applied to the determination of differential emission measures in the solar corona \citep{cheung2015}. Although minimizing the $\ell_1$-norm is not strictly equivalent to minimizing the number of non-zero components of $\xb$, the minimum-$\ell_1$ solution to an under-determined system of linear equations is often the sparsest solution to that system \citep{candes2006, donoho2008}. Numerically, it is preferable as it is a convex optimisation problem that can be efficiently solved using linear programming \citep{gill2011}. For the examples in Section \ref{sec:tests}, we used the numerical implementation of basis pursuit in the SparseLab\footnote{SparseLab is freely available for download from \url{http://sparselab.stanford.edu}.} library.  This implementation of basis pursuit uses a primal dual method \citep{chen2001} and has a \edit{single parameter controlling the error tolerance} for the optimization.

For all of the test cases in Sections \ref{sec:tests} and \ref{sec:limitations}, with tolerance $10^{-12}$, the optimization took only 10 to 20 seconds on a desktop workstation, not including the time to construct the matrix $A$ (which is only needed once for each grid). This is certainly fast enough for practical application to a time sequence of $\dbr$ maps, as would be needed for driving a coronal MHD simulation.

\begin{figure*}
\includegraphics[width=\textwidth]{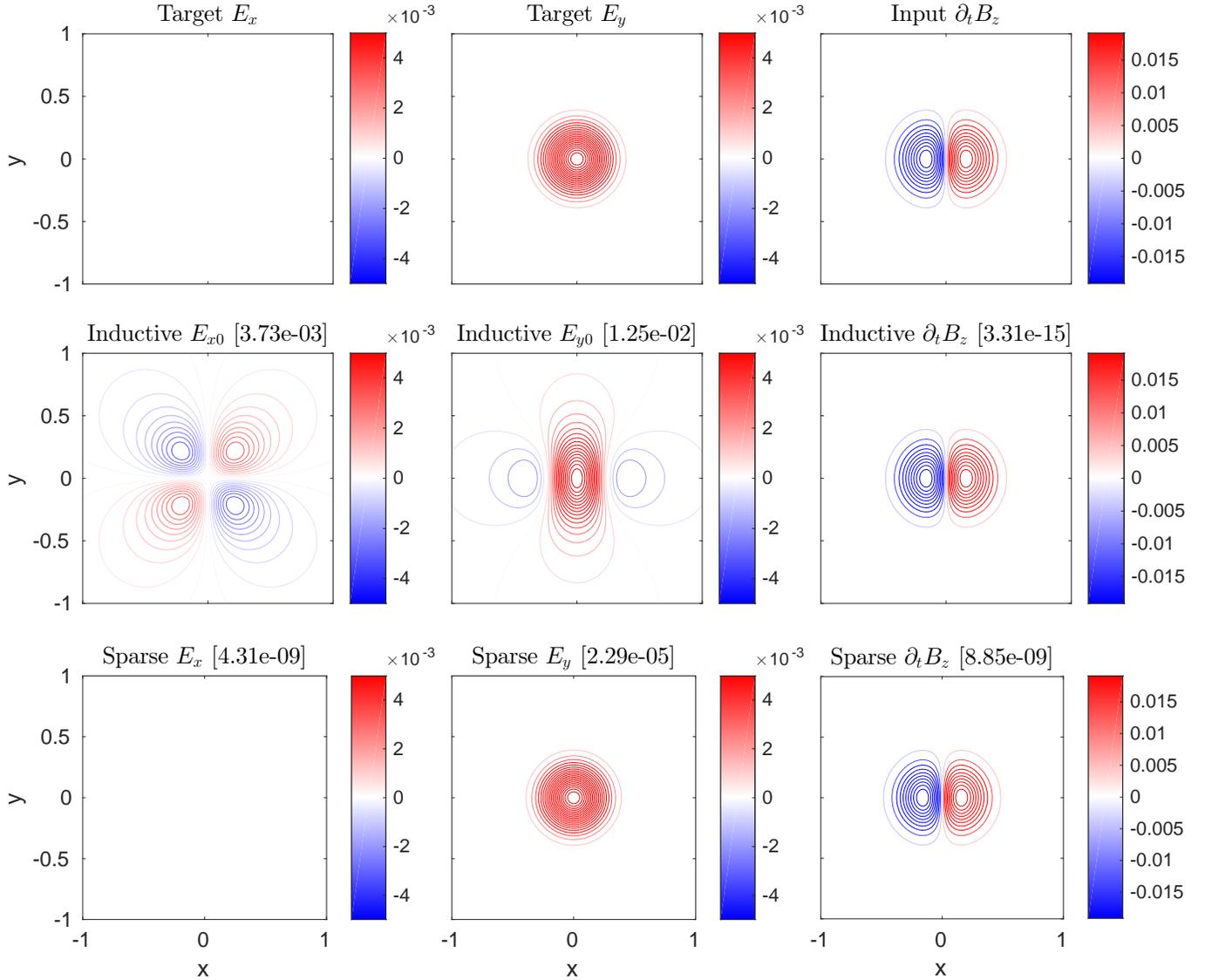}
\caption{Electric field reconstructions from (\ref{eqn:bca}) with $\delta^2=0.05$, discretized with grid size $n=256$ (only part of the domain is shown). The top row shows the target solution, middle row the inductive solution, and bottom row the sparse solution (with tolerance $10^{-12}$). The numbers in brackets are maximum absolute errors.}
\label{fig:caseA}
\end{figure*}

%%%%%%%%%%%%%%%%%%%%%%%
%%%%%%%%%%%%%%%%%%%%%%%
\section{Numerical Tests} \label{sec:tests}

In this section, we compare the inductive and sparse $\Ep$ solutions for two test cases. In the first case (Section \ref{sec:bip}), the true $\Ep$ is known through Ohm's law, while in the second (Section \ref{sec:composite}) it is not. Moreover, the first case uses Cartesian geometry and the second case uses spherical geometry. The additional problems that arise when applying the technique to real magnetic data are discussed in Section \ref{sec:limitations}.

\subsection{Bipolar distribution} \label{sec:bip}

Our first test case is based on the simplest configuration that may arise from Ohm's law. In fact, due to the linearity of both Ohm's law and Equation 
(\ref{eqn:faraday1}), more general test configurations may readily be obtained by superimposing multiple copies of this basic configuration. For simplicity, the computation is done in \edit{Cartesian} coordinates, in a square domain $|x|<3$, $|y|<3$, using the discretization described in Section \ref{sec:dis}. Periodic boundary conditions are applied in both directions.

Our basic solution is the bipolar distribution
\begin{equation}
\dbz = x\exp\left({-\frac{x^2 + y^2}{\delta^2}}\right).
\label{eqn:bca}
\end{equation}
Physically, this could arise from uniform advection of a single magnetic polarity $B_z = \exp[-(x^2 + y^2)/\delta^2]$ under an ideal Ohm's law $\Ep = -\vb_\perp\times(B_z\evz)$ with velocity $\vb_\perp = \delta^2/2\evx$, in which case
\begin{equation}
\Ep = \frac{\delta^2}{2}\exp\left({-\frac{x^2 + y^2}{\delta^2}}\right).
\label{eqn:eca}
\end{equation}
(In this case, we are taking a snapshot at $t=0$, as the polarity is centred on the origin.)
Alternatively, the same $\dbz$ (with different $B_z$) might correspond to the emergence of a bipolar magnetic field from the solar interior. The target $\Ep$ given by (\ref{eqn:eca}), along with the $\dbz$ in (\ref{eqn:bca}), are illustrated in the top row of Figure \ref{fig:caseA}.

\begin{figure}
\centering
\includegraphics[width=\columnwidth]{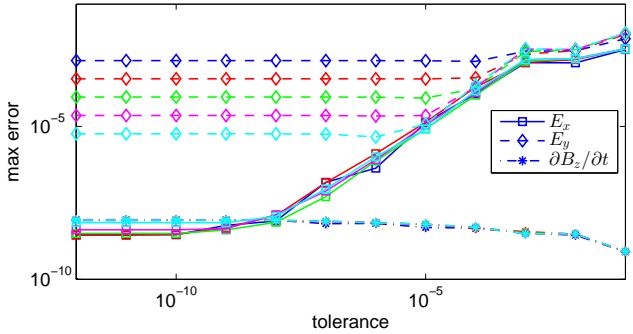}
\caption{Convergence of the sparse solution for (\ref{eqn:bca}), as a function of the tolerance parameter ($x$-axis) and grid resolution (colors). The maximum absolute errors are shown for $E_x$ (solid lines/squares), $E_y$ (dashed lines/diamonds) and $\dbr$ (dot-dashed lines/asterisks). The colors refer to grids with $n=32$ (blue), $n=64$ (red), $n=128$ (green), $n=256$ (magenta), and $n=512$ (cyan).}
\label{fig:convA}
\end{figure}

The middle row of Figure \ref{fig:caseA} shows the inductive solution $\Epo$ for this $\dbz$, which is qualitatively similar to the illustrative solution in Section \ref{sec:inductive}. Not only is $\Epo$ not localized within the region of non-zero $\dbz$, but the topology is wrong: there is a substantial $E_{x0}$ component, despite the fact that $E_x=0$ for the target solution. The sparse solution using basis pursuit, shown in the bottom row of Figure \ref{fig:caseA}, is substantially more accurate. To demonstrate convergence, Figure \ref{fig:convA} shows the absolute errors \edit{$\max|E_x - E_x^{\rm target}|$ and $\max|E_y-E_y^{\rm target}|$} for the sparse solution, as a function of both grid resolution and of the tolerance parameter in the basis pursuit algorithm. Firstly, the error in $\dbz$ is independent of tolerance or grid resolution, indicating that the constraint $A\xb=\bb$ is preserved to high accuracy in all cases. Secondly, there is convergence in both $E_x$ and $E_y$ as the tolerance is reduced. The error in $E_x$ is independent of resolution, whereas that in $E_y$ saturates at a (higher) level depending on grid resolution. This simply reflects the truncation error in the numerical approximation (\ref{eqn:discrete}) for the curl, which is quadratic in $\Delta x$. The saturation is not seen in $E_x$ because of our particular target solution $E_x=0$.

\subsection{Spherical distribution} \label{sec:composite}

\begin{figure}
\centering
\includegraphics[width=\columnwidth]{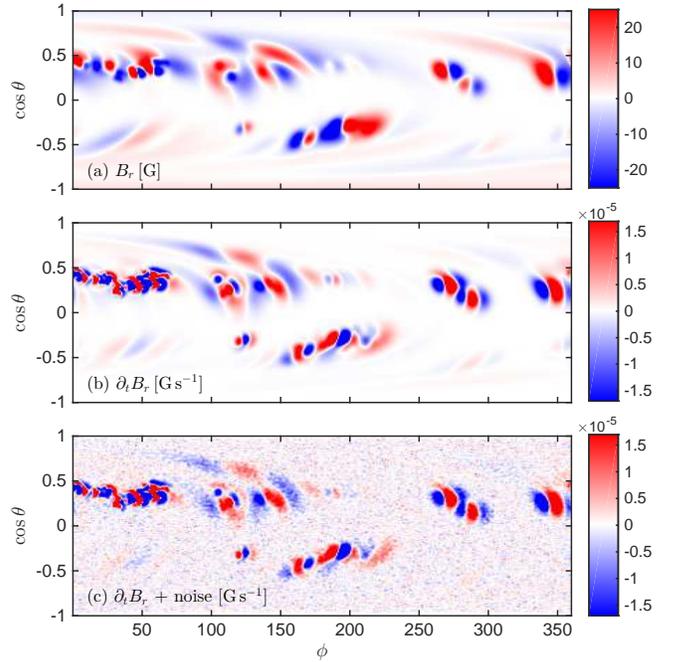}
\caption{Input data for the spherical test case, showing the distributions of (a) $B_r$ and (b) $\dbr$ from the flux transport model. Panel (c) shows the map of $\dbr$ with added noise that we use for the test. The maps are saturated at $\pm20\%$ of maximum for (a), and $\pm5\%$ of maximum for (b) and (c). }
\label{fig:brsft}
\end{figure}

\begin{figure*}
\centering
\includegraphics[width=\textwidth]{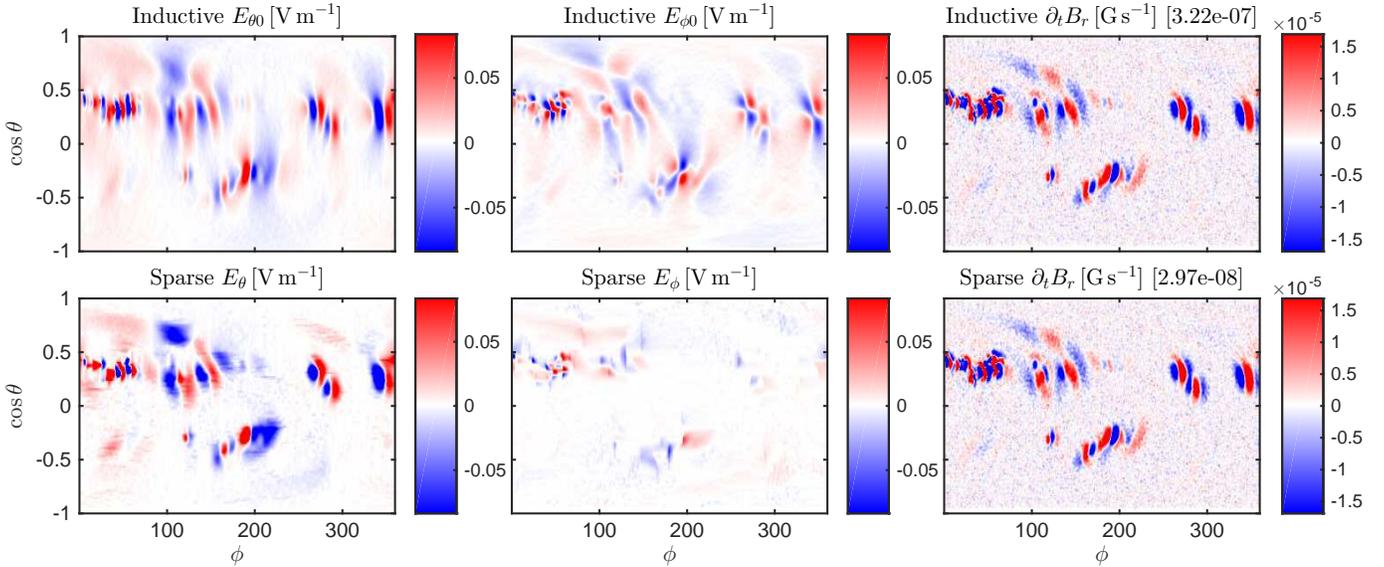}
\caption{Electric field reconstructions for the spherical test case (Figure \ref{fig:brsft}c). The top row shows the inductive solution and the bottom row the sparse solution (with tolerance $10^{-12}$). For comparison, all electric fields are saturated at $\pm20\%$ of the maximum inductive $|E_\theta|$, while $\dbr$ is saturated at $\pm5\%$ of the maximum input value. The numbers in brackets give maximum absolute (discretization) errors.}
\label{fig:esft}
\end{figure*}

To demonstrate how the sparse reconstruction performs on a more realistic example, we have taken a snapshot from a flux-transport simulation of $B_r(\theta,\phi,t)$ in spherical coordinates, covering the full solar surface. The reason for using a simulated map rather than an observed synoptic magnetogram is to ensure perfect flux balance; the consequences of not having flux balance will become evident in Section \ref{sec:limitations}.

We used the flux-transport model described by \citet{yeates2015}, for which the numerical code is freely available (\url{https://github.com/antyeates1983/sft_data}). As input data, we used synoptic maps of $B_r$ from the \textit{Global Oscillation Network Group} (GONG, \url{gong.nso.edu/data/magmap/}), starting in Carrington rotation 2073 and generating the output snapshot in rotation 2109. As described by \citet{yeates2015}, the input maps are used (a) to initialize $B_r$ at the beginning of the simulation, and (b) to extract strong flux regions used to update $B_r$ during the evolution. The computation used a $360\times180$ grid, equally spaced in longitude ($\phi$) and sine latitude ($\cos\theta$).

For context, Figure \ref{fig:brsft}(a) shows the $B_r$ map from the flux-transport simulation, while  Figure \ref{fig:brsft}(b) shows the $\dbr$ map from the same time. Since the simulation uses a supergranular diffusion rather than imposing convective velocities, the map appears smoother than would a real observed map. To make the $\dbr$ map look more realistic, we have added random noise to produce the map in Figure \ref{fig:brsft}(c). The noise is carefully constructed to preserve local flux balance, and is generated by adding a bipolar distribution
\begin{equation}
\beta^{i_0,j_0}(i-i_0)\exp\left(-\frac{(i-i_0)^2 + (j-j_0)^2}{\delta^{i_0,j_0}}\right)
\end{equation}
centred at each pixel $(i_0,j_0)$ on the $(s,\phi)$ grid. The magnitude $\beta^{i_0,j_0}$ at each pixel is chosen from a normal distribution with mean zero and $\sigma=0.005\max_{i_0,j_0}|\dbr|$, and the size $\delta^{i_0,j_0}$ is either 1 or 2 pixels, with equal probability. In addition, we rotate the pattern by 90 degrees at each particular pixel with equal probability. This generates a more realistic map as shown in Figure \ref{fig:brsft}(c).

Figure \ref{fig:esft} shows the inductive and sparse reconstructions of $\Ep$ from the map in Figure \ref{fig:brsft}(c). To account for the spherical coordinates, Equations (\ref{eqn:edis1})--(\ref{eqn:pdis}) have been modified to include the necessary coordinate factors. The right-hand column of Figure \ref{fig:esft} shows that both methods preserve $\dbr$ to high accuracy, as in the Cartesian test. Unlike in the Cartesian test, however, we no longer compare to a target $\Ep$, since the $\Ep$ corresponding to our added noise is unknown. Nevertheless, \edit{we can still see} that the sparse reconstruction is \edit{superior to} the inductive reconstruction. \edit{This is because, although} the $\dbr$ distribution is more complex than in the previous test, the sparse $\Ep$ is still much \edit{better} localized within regions of strong $\dbr$ \edit{than is the inductive $\Ep$}. Moreover, it has a weaker $E_\phi$ than $E_\theta$, consistent with the fact that the dominant Ohm's law contribution in the flux-transport simulation was from differential rotation ($v_\phi$). This is not the case in the inductive solution. Thus we are confident that the sparse solution \edit{is superior to the inductive solution} not just for simple test cases, but also for more complex $\dbr$ distributions that are qualitatively similar to those on the Sun.

%%%%%%%%%%%%%%%%%%%%%%%
%%%%%%%%%%%%%%%%%%%%%%%
\section{Limitations} \label{sec:limitations}

As with any inverse problem, care is needed in the application of our sparse reconstruction technique. In this section, we consider two limitations which can be important in the practical application to the driving of coronal MHD models.

\subsection{Diffuse distributions of $\dbr$} \label{sec:diffuse}

The sparse solution makes the physical assumption that $\Ep$ should be localized because it satisfies Ohm's law for a localized $\Bb$. If the magnetic field distribution is too diffuse, then we can not expect the assumption of localization to recover the correct (diffuse) $\Ep$.

\begin{figure*}
\includegraphics[width=\textwidth]{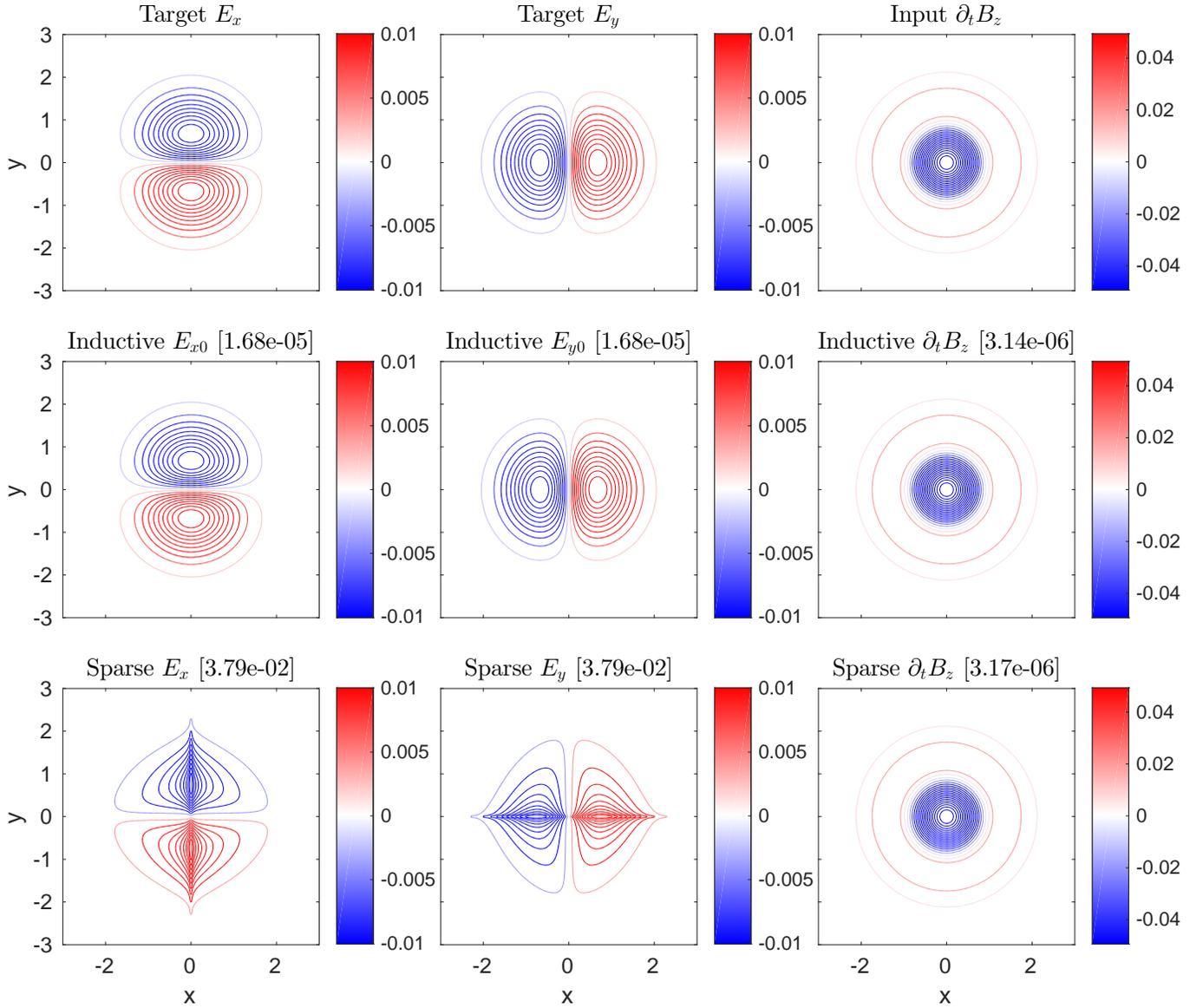}
\caption{Electric field reconstructions from (\ref{eqn:bcb}), discretized with grid size $n=256$. The top row shows the target solution, middle row the inductive solution, and bottom row the sparse solution (with tolerance $10^{-12}$). The numbers in brackets are maximum absolute errors.}
\label{fig:caseB}
\end{figure*}

To illustrate what happens if we nevertheless try to apply the sparse reconstruction in such a situation, consider the horizontal diffusive spreading of an initial gaussian magnetic polarity
\begin{equation}
B_z(x,y,t) = \frac{a^2}{a^2 + 4\eta t}\exp\left(-\frac{x^2 + y^2}{a^2 + 4\eta t}\right).
\end{equation}
From the resistive Ohm's law $\Ep = \eta\nabla\times(B_z\evz)$, we get
\begin{eqnarray}
E_x = -\frac{2a^2 y}{(a^2 + 4\eta t)^2} \exp\left(-\frac{x^2 + y^2}{a^2 + 4\eta t}\right), \\
E_y = \frac{2a^2 x}{(a^2 + 4\eta t)^2} \exp\left(-\frac{x^2 + y^2}{a^2 + 4\eta t}\right).
\end{eqnarray}
Faraday's law then gives
\begin{eqnarray}
\dbz = 4a^2\eta\left[\frac{x^2 + y^2}{(a^2 + 4\eta t)^3} - \frac{1}{(a^2 + 4\eta t)^2}\right]\nonumber\\
\times\exp\left(-\frac{x^2 + y^2}{a^2 + 4\eta t}\right).
\label{eqn:bcb}
\end{eqnarray}
For this exercise, we fix $a=0.5$ and $\eta t = 0.1$. This solution is shown in the top row of Figure \ref{fig:caseB}.

It is important to note that, in this case, Ohm's law is compatible with a purely inductive solution, because $\Ep$ from Ohm's law has the inductive form with $\Phi(x,y,t) = \eta B_z(x,y,t)$. Indeed, the middle row of Figure \ref{fig:caseB} shows the inductive solution to reproduce the target $\Ep$ in this case. However, the bottom row shows that the sparse solution using basis pursuit does not converge to the target. Rather, it favors a concentration of $E_x$ along vertical lines, and $E_y$ along horizontal lines. This behavior is typical when the target $\Ep$ is too diffuse. It is not possible for the total spatial extent of $\Ep$ to be more localized than $\dbr$, owing to the constraint of satisfying (\ref{eqn:faraday1}). But it \emph{is} possible for more of the electric field to be concentrated in thinner regions, and this is what minimizing the $\ell_1$-norm does. 

This limitation is an obvious one, and limits the sparse solution technique to input maps where $\Bb$ is sufficiently localized. Fortunately, this situation is typical on the Sun, given high enough resolution. The flux-transport test case in Section \ref{sec:composite} showed the method to work for realistic solar magnetic maps, at the resolutions of present-day simulations. However, there is another more subtle problem with real data, that we describe in the following section.

\subsection{Flux imbalance in $\dbr$ maps}

\begin{figure*}
\includegraphics[width=\textwidth]{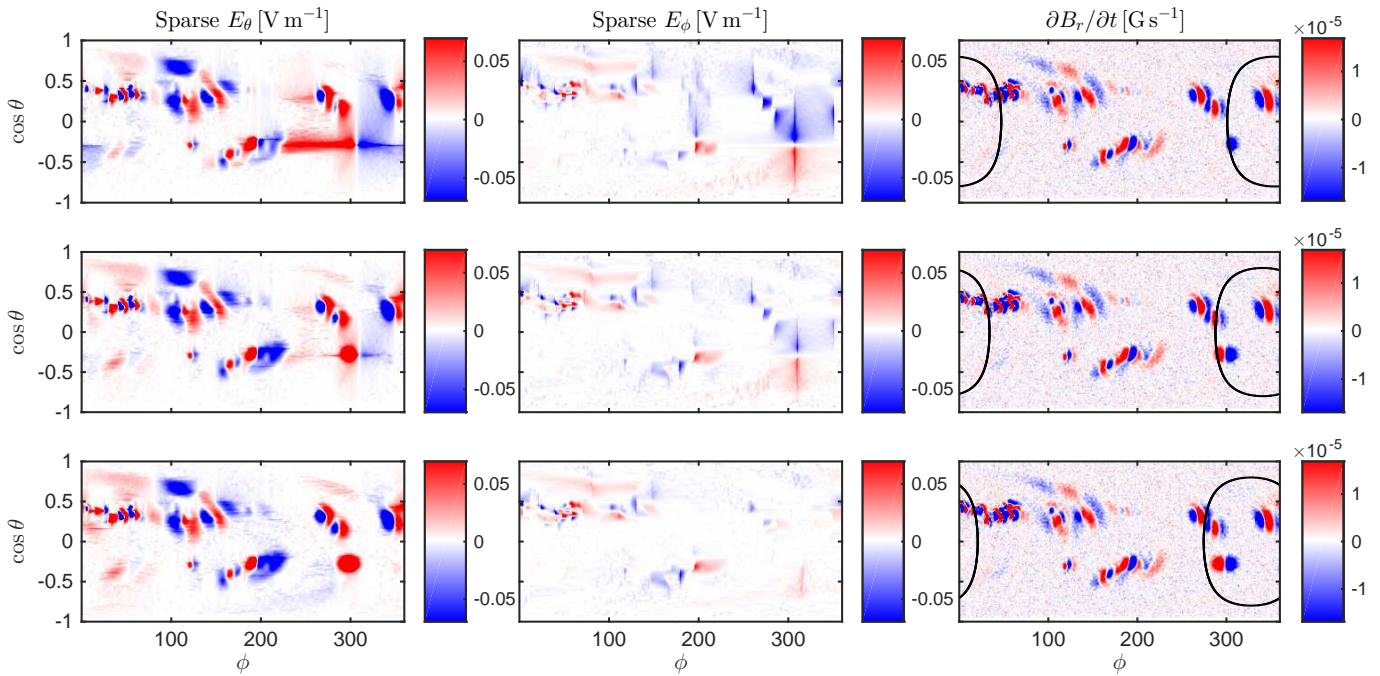}
\caption{Sparse electric field reconstructions from the spherical test case (Figure \ref{fig:brsft}c), modified to simulate assimilation of a new bipolar region. Each row simulates a different time of observation where the assimilation window (black outline in the right-hand column) has shifted leftward by one day of solar rotation. Here no correction for the flux imbalance in $\dbr$ has been applied.}
\label{fig:imbalance}
\end{figure*}

For a localized $\Ep$ to exist, it is necessary not only that $\dbr$ is localized, but that the net flux $\int_S\dbr\,\mathrm{d}A$ vanishes over the region $S$ of localization. To see this, apply Stokes' theorem on some closed curve that encircles $S$. If the net flux in $S$ is non-zero, then there must be non-zero $\Ep$ somewhere on the bounding curve, and indeed on any curve enclosing non-zero net flux.

This condition of vanishing net flux is satisfied by both of our simple examples (\ref{eqn:bca}) and (\ref{eqn:bcb}), and by every local flux concentration in the flux-transport test case (Section \ref{sec:composite}). But it need not be satisfied in an observationally-derived map of $\dbr$, even if a flux correction has been applied. We will demonstrate this problem firstly in a controlled test and secondly in a ``real'' data-assimilative model.

For the controlled test, we modify the input map from Section \ref{sec:composite} to simulate the typical situation where new magnetogram observations are assimilated only within a finite region on the visible face of the Sun. The most severe problems of flux imbalance occur when a new active region has emerged on the far-side of the Sun, and is gradually assimilated into the magnetic map as it rotates into view.

Figure \ref{fig:imbalance} shows our test. Here the original $\dbr$ map, from Figure \ref{fig:brsft}(c), has been modified to simulate the gradual assimilation of a new bipolar active region. In the right-hand column of Figure \ref{fig:imbalance}, the assimilation region is indicated on the modified $\dbr$ maps. Each row corresponds to a different time as the new active region rotates into view, with the times differing by one day of solar rotation ($27.2753^\circ$ in these Carrington maps).

When the new region is fully contained within the assimilation window (bottom row of Figure \ref{fig:imbalance}), it makes only a localized contribution to the sparse $\Ep$, correctly reproducing $\Ep$ as in Section \ref{sec:bip}. But when the region is only partially observed, there is an unbalanced flux of $\dbr$, both in the assimilation window and in the entire map. Correspondingly, it is impossible to find a localized electric field, and indeed the sparse solution fails. In the area of the new region, we see similar linear structures to the diffusive example in Figure \ref{fig:caseB}, where the flux was also not locally balanced. But notice that the electric field is modified not only near to the new region, but also at farther distances.

\begin{figure*}
\includegraphics[width=\textwidth]{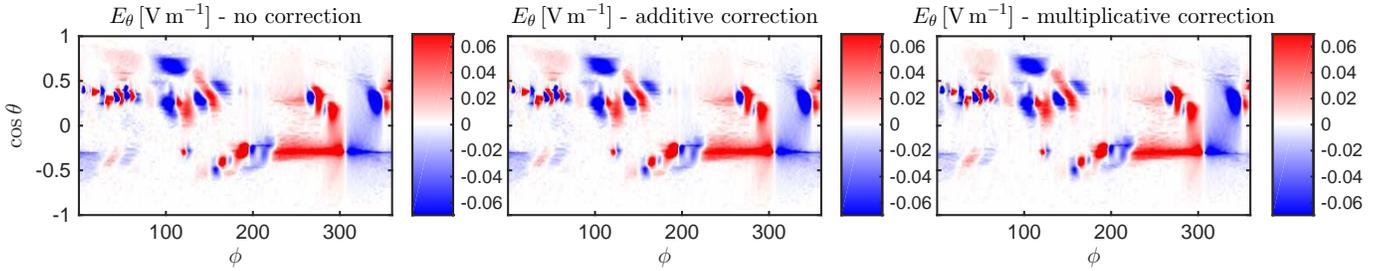}
\caption{Sparse $E_\theta$ for different methods of global flux correction.}
\label{fig:balance}
\end{figure*}

Of course, this is a rather unrealistic test. If this $\Ep$ was being used to drive a coronal MHD model, for example, one would correct the $\dbr$ map for flux balance, before trying to compute $\Ep$. In Figure \ref{fig:balance} we show the effect on $E_\theta$ of first making a flux-balance correction to the $\dbr$ map (from the top row of Figure \ref{fig:imbalance}). The middle and right plots show the results with two different methods of flux correction. In the ``additive'' method, an equal amount is subtracted from each pixel in the $360\times 180$ grid to remove the imbalance. In the ``multiplicative'' method, all of the pixels where $\dbr > 0$ are multiplied by a constant factor $f_+$, and all pixels where $\dbr < 0$ are multiplied by another constant factor $f_-$. The factors $f_+$, $f_-$ are chosen so that the net flux vanishes after scaling, and is equal to the mean of the positive and negative fluxes before scaling. Comparing the results in Figure \ref{fig:imbalance}, it is evident that neither method of flux correction gives much improvement in the reconstructed $E_\theta$ (or $E_\phi$). Only by limiting the flux correction to the active region itself could the solution be improved.

Similar behavior is found when we apply the sparse reconstruction technique to a time sequence of $B_r$ maps generated by the Air Force Data-Assimilative Photospheric Flux Transport (ADAPT) model \citep{arge2010,henney2012}. This model assimilates observed magnetograms from the visible face of the Sun into a surface flux-transport simulation, so as to approximate the global distribution of $B_r$ on the solar surface, as a function of time. The data assimilation means that the electric field $\Ep$ is not known everywhere in the map, so must be reconstructed, if these maps are used to drive coronal MHD simulations.

\begin{figure}
\centering
\includegraphics[width=\columnwidth]{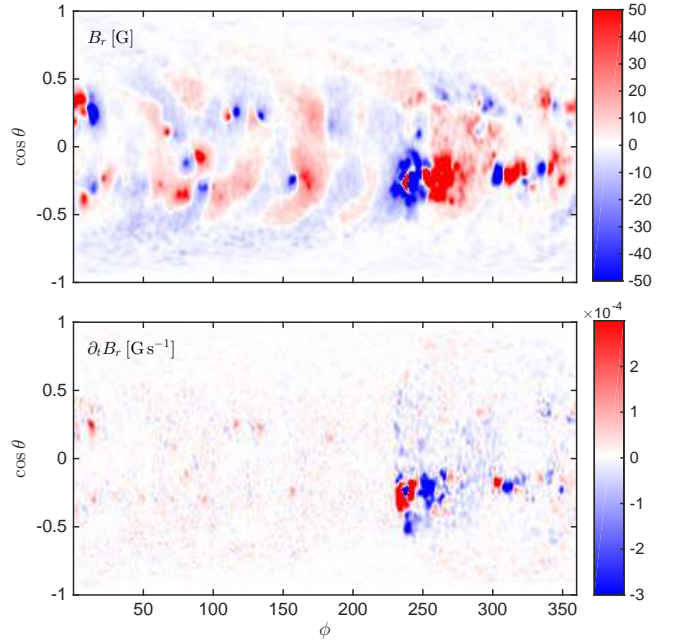}
\caption{Input from the ADAPT model, showing $B_r$ (for context) and the computed $\dbr$. The plots have been saturated at $\pm 50\,\mathrm{G}$ and $\pm 3\times 10^{-4}\,\mathrm{G}\,\mathrm{s}^{-1}$, respectively.}
\label{fig:bradapt}
\end{figure}

As an illustration, we choose an ADAPT run driven by GONG magnetograms, and extract $\dbr$ for 2014 November 16, 00:00 UT. This particular date has been deliberately chosen since  \citet{weinzierl2016} found a significant non-localized $\Epo$ caused by re-assimilation of a large active region (see their Figure 11). Here, we process the ADAPT data by (i) applying a multiplicative correction for flux balance; (ii) applying a spatial smoothing to the maps; and (iii) remapping to a $360\times 180$ grid in $\phi$ and $\cos\theta$. The time derivative $\dbr$ is then estimated using a Savitzky-Golay smoothing filter (of total width 18 hrs), generating the map shown in Figure \ref{fig:bradapt}. The position of the data-assimilation window is evident in the map of $\dbr$, with the largest contribution coming from the main active region AR12209. (This is not newly emerging, but has significantly changed its structure since it was last observed on the previous Carrington rotation.)

\begin{figure*}
\includegraphics[width=\textwidth]{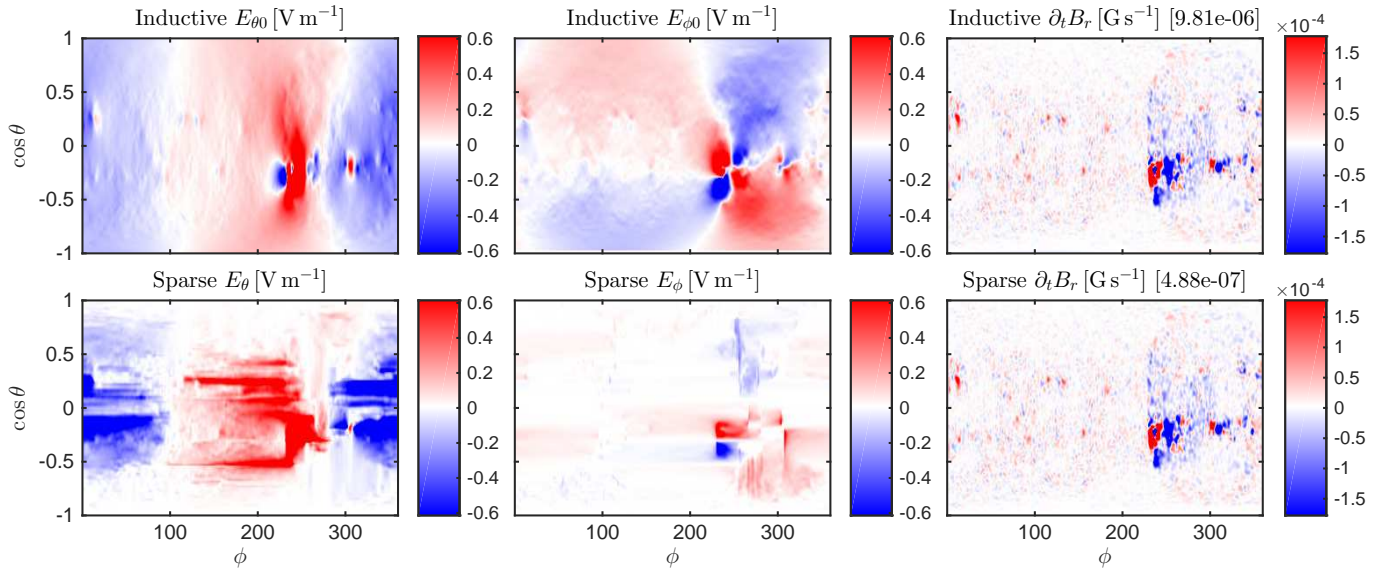}
\caption{Inductive (top) and sparse (bottom) electric field solutions, for the $\dbr$ map derived from ADAPT $B_r$ maps. The electric field components are saturated at $\pm 20\%$ and $\dbr$ at $\pm10\%$.}
\label{fig:eadapt}
\end{figure*}

In Figure \ref{fig:eadapt}, we show the inductive and sparse solutions for $\Ep$ in this case. Since there is such a dominant source of $\dbr$, there are significant non-local electric fields in the inductive solution, in contravention of Ohm's law. The sparse $E_\phi$ is rather better, but the sparse $E_\theta$ is still not correctly localized. The cause of this is the flux imbalance in the original ADAPT map for $\dbr$, as in the previous test (Figure \ref{fig:imbalance}). Again the global flux correction has not really helped the problem. We conclude from these tests that lack of local flux-balance is the main obstacle to driving coronal MHD models with data-assimilative magnetic maps.

\section{Conclusion} \label{sec:conc}

In this paper, we have shown how a sparse reconstruction technique based on the idea of searching for a localized solution allows one to recover accurate Ohm's law $\Ep$ based on seemingly insufficient data of just $\dbr$. The technique is potentially useful for driving coronal MHD simulations from time-sequences of photospheric $B_r$ maps.

However, we have also identified a difficulty that can arise if one tries to compute the sparse electric field from maps where observational magnetogram data have been assimilated \citep[e.g.,][]{schrijver2003,upton2014,hickmann2015}. Namely, errors in $\dbr$ that lead to local flux imbalance can prevent the sparse solution from being a good approximation to the real $\Ep$ (meaning the $\Ep$ that satisfies both Faraday's law and Ohm's law). This implies that the $\dbr$ map must first be corrected if $\Ep$ is to be reconstructed successfully. We have shown in Figure \ref{fig:balance} that straightforward ``global'' methods of correcting the flux -- whether additive or multiplicative -- are insufficient to remove the problem. But this does not mean that a more sophisticated pre-processing method could not be successful; this is a possible direction for future research.

The problem of flux balance is probably best addressed at source, when the $B_r$ maps themselves are produced. This is easy to achieve in an idealized flux-transport model, but not in one with direct assimilation of observed magnetogram data. \citet{yeates2015} have introduced a flux-transport model where entire, flux-balanced, bipolar regions are assimilated based on synoptic magnetograms (as used for our test case in Section \ref{sec:composite}). But it remains an open problem to assimilate higher-resolution magnetogram data in a manner that enforces, as far as possible, local flux-balance in $B_r$.

\acknowledgments

The author is grateful to Lockheed Martin and UC Berkeley for hosting my research visit in 2015, and in particular to Mark Cheung for suggesting the idea of looking for sparse solutions. I also thank Zoran Miki\'c and Cooper Downs (Predictive Science Inc.), and Roger Fletcher (University of Dundee, now sadly deceased) for useful discussions. The work was supported by a grant from the US Air Force Office of Scientific Research (AFOSR) in the AFOSR Basic Research Initiative “Understanding the Interaction of CMEs with the Solar-Terrestrial Environment.” I thank Carl Henney for supplying the ADAPT maps. This work utilises data obtained by the Global Oscillation Network Group (GONG) program, managed by the National Solar Observatory, which is operated by AURA, Inc. under a cooperative agreement with the National Science Foundation. The data were acquired by instruments operated by the Big Bear Solar Observatory, High Altitude Observatory, Learmonth Solar Observatory, Udaipur Solar Observatory, Instituto de Astrofísica de Canarias, and Cerro Tololo Interamerican Observatory.

\bibliographystyle{aasjournal}
\bibliography{yeates_electric}


\listofchanges

\end{document}